\documentclass[%
 reprint,
 amsmath,amssymb,
 aps,
]{revtex4-2}

\usepackage{graphicx}% Include figure files
\usepackage{dcolumn}% Align table columns on decimal point
\usepackage{bm}% bold math
\usepackage{babel}

\usepackage{comment}
\usepackage[svgnames]{xcolor}
\usepackage{xstring}
\newcommand{\EatOneArg}[1]{}

\usepackage{graphicx}% Include figure files
\usepackage{dcolumn}% Align table columns on decimal point
\usepackage{bm}% bold math
\usepackage[hidelinks]{hyperref}% add hypertext capabilities

\usepackage{xcolor}

\newcommand{\diff}{\mathrm{d}}

\newcommand{\bx}{\mathbf{x}}
\newcommand{\bk}{\mathbf{k}}

\newcommand{\oma}{\omega}

\newcommand{\cg}{c_g}

\newcommand{\Za}{Z_\alpha}
\newcommand{\Zb}{Z_\beta}

\newcommand{\na}{n_\alpha}
\newcommand{\nb}{n_\beta}
\newcommand{\nc}{n_\gamma}
\renewcommand{\oma}{\omega_\alpha}
\newcommand{\omb}{\omega_\beta}
\newcommand{\omg}{\omega_\gamma}

\newcommand{\mean}[1]{\overline{#1}}

\renewcommand{\Re}{\operatorname{Re}}

\begin{document}

%lyx shortcuts
\global\long\def\dyad#1{\underline{\underline{\boldsymbol{#1}}}}%
\global\long\def\ubar#1{\underbar{\ensuremath{\boldsymbol{#1}}}}%
\global\long\def\integer{\mathbb{Z}}%
\global\long\def\natural{\mathbb{N}}%
\global\long\def\real#1{\mathbb{R}^{#1}}%
\global\long\def\complex#1{\mathbb{C}^{#1}}%
\global\long\def\defined{\triangleq}%
\global\long\def\trace{\text{trace}}%
\global\long\def\del{\nabla}%
\global\long\def\cross{\times}%
\global\long\def\diff#1#2{\frac{\partial#1}{\partial#2}}%
\global\long\def\Diff#1#2{\frac{d#1}{d#2}}%
\global\long\def\bra#1{\left\langle #1\right|}%
\global\long\def\ket#1{\left|#1\right\rangle }%
\global\long\def\braket#1#2{\left\langle #1|#2\right\rangle }%
\global\long\def\ketbra#1#2{\left|#1\right\rangle \left\langle #2\right|}%
\global\long\def\identity{\mathbf{1}}%
\global\long\def\paulix{\begin{pmatrix}  &  1\\
 1 
\end{pmatrix}}%
\global\long\def\pauliy{\begin{pmatrix}  &  -i\\
 i 
\end{pmatrix}}%
\global\long\def\pauliz{\begin{pmatrix}1\\
  &  -1 
\end{pmatrix}}%
\global\long\def\sinc{\mbox{sinc}}%
\global\long\def\four{\mathcal{F}}%
\global\long\def\dag{^{\dagger}}%
\global\long\def\norm#1{\left\Vert #1\right\Vert }%
\global\long\def\hamil{\mathcal{H}}%
\global\long\def\tens{\otimes}%
\global\long\def\ord#1{\mathcal{O}\left(#1\right)}%
\global\long\def\undercom#1#2{\underset{_{#2}}{\underbrace{#1}}}%
 
\global\long\def\conv#1#2{\underset{_{#1\rightarrow#2}}{\longrightarrow}}%
\global\long\def\tg{^{\prime}}%
\global\long\def\ttg{^{\prime\prime}}%
\global\long\def\clop#1{\left[#1\right)}%
\global\long\def\opcl#1{\left(#1\right]}%
\global\long\def\broket#1#2#3{\bra{#1}#2\ket{#3}}%
\global\long\def\div{\del\cdot}%
\global\long\def\rot{\del\cross}%
\global\long\def\up{\uparrow}%
\global\long\def\down{\downarrow}%
\global\long\def\Tr{\mbox{Tr}}%

\global\long\def\per{\mbox{}}%
\global\long\def\pd{\mbox{}}%
\global\long\def\p{\mbox{}}%
\global\long\def\ad{\mbox{}}%
\global\long\def\a{\mbox{}}%
\global\long\def\la{\mbox{\ensuremath{\mathcal{L}}}}%
\global\long\def\cm{\mathcal{M}}%
\global\long\def\cg{\mbox{\ensuremath{\mathcal{G}}}}%
%end lyx shortcuts

\preprint{APS/123-QED}

\title{The role of sign indefinite invariants in shaping %geophysical 
turbulent %energy 
cascades}

\author{Michal Shavit, Oliver B\"uhler 
and Jalal Shatah}

\affiliation{%
 Courant Institute of Mathematical Sciences, New York University, NY 10012, USA.
}%


\begin{abstract}
%needs to be < 600 characters 
We highlight a non-canonical yet natural choice of variables for an efficient derivation of a kinetic equation for the energy density in non-isotropic systems, including internal gravity waves on a vertical plane, inertial and Rossby waves. The existence of a second quadratic invariant %, regardless of its sign, 
simplifies the kinetic equation and leads to extra conservation laws for resonant interactions. We analytically determine the scaling of the radial %, integrated over the angle,   
turbulent energy spectrum. Our findings suggest the existence of an inverse energy cascade of internal gravity waves, from small to large scales, in practically relevant scenarios.
\\

\end{abstract}

\maketitle

%\tableofcontents

%\section{\label{sec:level1}Introduction}
\textbf{\textit{Introduction}}
Strongly dispersive waves are ubiquitous in geophysical fluid dynamics, where they occur on scales from centimeters to thousands of kilometres and contribute in an essential and intricate way to the long-term nonlinear dynamics of the climate system \cite{andrews1987middle,buhler2014waves,vallis2017atmospheric,whalen2020internal}.  Examples include surface waves, internal inertia--gravity waves, and Rossby waves, all of which owe their existence to some combination of gravity, rotation, and curvature of the Earth.  Many of these waves are far too small in scale to be resolvable numerically, making their study a pressing issue for theoretical modeling and investigation.  For small-amplitude waves the methods of wave turbulence theory can play an important  part in this, because they produce  a closed kinetic equation for the slow evolution of the averaged spectral energy density.   There has been significant progress for idealized model systems \citep[e.g.][]{ZLF,buckmaster}, but so far this has not yet been translated to systems of direct geophysical interest.   Arguably, progress has been hampered by the extremely cumbersome form taken by the relevant equations when attempting to shoe-horn them into classical wave turbulence theory, which was formulated in canonical variables for Hamiltonian systems \cite{zakharov1987hamiltonian,Lvovhamil}.  But the underlying fluid equations are non-canonical Hamiltonian systems, as is made obvious by the fact that the Euler equations are highly nonlinear yet  their energy function is quadratic \cite{shepherd90sym,salmon1998lectures}.  This has motivated the present work, in which we pursue a  reformulation of kinetic wave theory for a number of two-dimensional fluid systems with quadratic energies based on a particular choice of non-canonical variables.     The practical utility of our choice of variables, which was introduced  in a different context by \cite{ripa}, derives from  the existence of a second quadratic invariant in these systems, which, albeit not sign-definite, greatly simplifies the wave interaction equations.  We leverage these simplifications into a derivation of scaling laws for the isotropic component of wave spectra and we present evidence for the importance of these second invariants in shaping the overall wave spectra in certain situations.   \textit{Mutatis mutandis,} much of our analysis applies to waves in plasmas as well. 

The two-dimensional Boussinesq equations restricted to a vertical $xz$-plane can be written as 
\begin{align}\label{eq:2D Boussinesq}
       \Delta\psi_{t}+\left\{ \psi,\Delta\psi\right\} &=-N^{2}\eta_{x}\\ \eta_{t}+\left\{ \psi,\eta\right\} &=\psi_x.\nonumber
\end{align}
Here $z$ is the vertical and $x$ is the  horizontal coordinate with corresponding velocities $w$ and $u$, 
$\psi$ is a stream function such that $(\psi_x,\psi_z)=(w,-u)$ and $-\Delta \psi$ is the vorticity, $\eta$ is the vertical displacement, $N$ the constant buoyancy frequency and $\left\{ g,f\right\}=\partial_xg\partial_zf-\partial_zg\partial_xf$. The vertical buoyancy force $b=-N^2\eta$ opposes vertical displacements and derives from a consideration of potential energy  in the presence of gravity and non-uniform density.  It is easily checked that this system has two exact quadratic invariants: the total energy $E=\int d\mathbf{x} (-\psi\Delta \psi+N^2\eta^2)$ and the pseudomomentum $P=\int d\mathbf{x} \,\eta \Delta \psi$.  The subtleties associated with the Hamiltonian point of view of these equations can be appreciated by investigating the origin of these conservation laws by rewriting  \eqref{eq:2D Boussinesq} as
\begin{equation}\label{eq:e general eqn}
    \partial_{t}D\phi = \mathcal{J}\frac{\delta E}{\delta\left(D\phi\right)}. %+\mathcal{L}\phi.
\end{equation}
Here $\phi^T\left(\mathbf{x},t\right)=\left(\psi,\eta \right)$, $D=\text{diag}(-\Delta,N^2)$ is a Hermitian semi-positive-definite operator, and  
\begin{equation}\label{eq:J}
\mathcal{J(\phi)}=\frac{1}{2}\begin{pmatrix}\left\{ -\Delta\psi,\cdot\right\}  & \left\{ N^2\eta,\cdot\right\} +N^2\partial_x\\
\left\{ N^2\eta,\cdot\right\} +N^2\partial_x & 0
\end{pmatrix}%, \,\,+ \frac{N^{2}}{2}\begin{pmatrix}0 & 1\\
%1 & 0
%\end{pmatrix}\partial_{x}
\end{equation}
is a skew-symmetric operator representing the Poisson structure. 
%\addOB{(Are we sure about the sign of the top right nonlinear component of J?  It seems not be skew on arbitrary pairs of functions that way).  Of course, the term does not matter at all in the equation of motion.}
This is a non-canonical Hamiltonian system for the variables $D\phi$ based on the inner product
Hamiltonian function
\begin{equation}\label{eq:energy}
    E=\braket{\phi}{D\phi}  = \int\! d\mathbf{x}\, \phi^T D\phi.
\end{equation}
Energy conservation is then transparently linked to the time translation symmetry of $\mathcal{J}(\phi)$.
The pseudomomentum can be written as
\begin{equation}\label{eq:pm}
P=\braket{D\phi}{C D \phi}\quad\mbox{with}\quad
C= -\frac{1}{2N^2}\begin{pmatrix}0 & 1\\
1 & 0
\end{pmatrix}. %\!\int d\mathbf{x}\phi^{\dagger}D^2\phi,
\end{equation}
Therefore ${\delta P}/{\delta(D\phi)}=2CD\phi$ and 
\begin{equation}
    \mathcal{J}(\phi) \frac{\delta P}{\delta(D\phi)} = -\frac{(D\phi)_x}{2},
\end{equation}
which ensures the invariance of $P$ based on the $x$-translation symmetry of (\ref{eq:energy}). This suggests interpreting $P$ as a canonical horizontal momentum, even though it does not agree with the horizontal momentum of the fluid.  
Actually, $CD\phi$ is in the kernel of the nonlinear part of $\mathcal{J}$, 
which suggests interpreting  $P$ as  a Casimir of that Poisson degenerate structure \cite{olver, shepherd90sym}.
This degeneracy translates to gauge invariance in terms of Lagrangian coordinates \cite{olver1993applications}. Thus, the conservation of $P$ appears Casimir-like based on the nonlinear dynamics, but momentum-like based on the linear dynamics.  Calling $P$ the `pseudomomentum' is in accordance with established usage in geophysical fluid dynamics \citep[e.g. \S4.3 in][]{shepherd90sym} and wave--mean interaction theory \cite{buhler2014waves}.  So, whilst the conceptual origins of the two conservation laws for $E$ and $P$ are subtle and subject to interpretation, their actual functional expressions as quadratic forms $E=\braket{\phi}{D\phi}$ and $P=\braket{D\phi}{C D \phi}$ are completely straightforward, and generalize easily.  Following \cite{ripa}, we exploit this by expanding the flow in variables that diagonalize both $E$ and $P$.

\textbf{\textit{Wave mode expansion.}} We consider a periodic domain $\mathbf{x}\in \left[0,L\right]^2$ and expand $\phi(\bx,t) = \sum_\alpha Z_{\alpha}(t)e_{\alpha}(\bx)$ in terms of linear  wave modes, where $Z_\alpha(t)$ are complex scalar wave amplitudes and the $e_\alpha(\bx)$ are eigenvector functions for the linear part of (\ref{eq:2D Boussinesq}), i.e.,
\begin{align}
-i\omega_\alpha D e_{\alpha} &= {N^2}\begin{pmatrix}0 & 1\\
1 & 0
\end{pmatrix}\partial_x e_{\alpha}.\label{eq:lin eigen}
\end{align}
The $e_\alpha(\bx)$ are proportional to Fourier modes $\exp(i\bk\cdot\bx)$ 
with $\bk=(k_x,k_z)\in (2\pi\mathbb{Z}/L)^2$. 
If $\bk=K(\cos\theta,\sin\theta)$ then the dispersion relation is $\omega=\pm N\cos\theta$.  The multi-index $\alpha=(\sigma,\mathbf{k})$ combines branch choice $\sigma = \pm 1$ and wave number $\bk$ such that
\begin{equation}\label{eq:omega}
\omega_{\alpha=(\sigma,\bk)}=\sigma N \cos\theta_k. 
\end{equation}
With this convention the choice $\sigma=\pm1$ corresponds to right-going or left-going waves, respectively, which are vital physical characteristics.  The reality of $\phi$ implies that $Z_{(\sigma,\bk)}(t) = Z^*_{(\sigma,-\bk)}(t)$,
where the star denotes complex conjugation.  This holds separately within each branch.
The expansion diagonalizes the energy   
\begin{align}
%E & =\int\! d\alpha z_{\alpha}z_{\alpha}^{*},
E  = \sum _\alpha E_\alpha= \sum _\alpha Z_{\alpha}Z_{\alpha}^{*}\label{eq:energy z}
%E & =\epsilon^2 \sum _\alpha Z_{\alpha}Z_{\alpha}^{*},\label{eq:energy z}
\end{align}
and yields the exact equations 
\begin{align}\label{eq:z}
Z_{\alpha,t}+i\omega_{\alpha}Z_{\alpha} & = \sum_{ \beta, \gamma} \frac{1}{2}V_{\alpha}^{\beta\gamma}Z_{\beta}^{*}Z_{\gamma}^{*}.
\end{align}
The interaction coefficients $V_{\alpha}^{\beta\gamma}=
\braket{e_{\alpha}^*}{\mathcal{J}(e_\gamma)e_\beta+\mathcal{J}(e_\beta)e_\gamma}$ 
%\begin{align} V_{\alpha}^{\beta\gamma}&= \frac{1}{2}\braket{e_{\alpha}}{(\mathcal{J}(e_\gamma)e_\beta+\mathcal{J}(e_\beta)e_\gamma)^*} \end{align}  
 are real, symmetric in upper indices %,$\sigma_\alpha^{\beta \gamma}=\sigma_\alpha^{\gamma \beta}$  
and zero unless $\bk_\alpha+\bk_\beta+\bk_\gamma = 0$. The expansion also diagonalizes the pseudomomentum 
\begin{align} \label{eq:PM z}
P  = \sum _\alpha P_\alpha= \sum _\alpha s_\alpha Z_{\alpha}Z_{\alpha}^{*},
\end{align}
where the horizontal slowness $s_\alpha=k_x/\oma$.  Hence $P_\alpha=k_x/\omega_\alpha \,E_\alpha$ has the sign of the horizontal phase (or group) velocity.

\textbf{\textit{The kinetic equation.}} From \eqref{eq:energy z} and \eqref{eq:z} the road to a kinetic equation is short. %\eqref{eq:z} is written in terms of small amplitude waves, where the explicit small parameter $0<\eps\ll1$ makes $\Za=O(1)$. %Setting $\epsilon=1$, 
%Assuming small initial conditions, we expand the field  $\phi$ in terms of small amplitude waves $\phi= \epsilon \sum_\alpha Z_{\alpha}e_{\alpha}$, where the explicit small parameter $0<\eps\ll1$ makes $\Za=O(1)$ %Such small uniform parameter relevant for internal wave interaction is the root mean square vertical gradient of the vertical displacement
The modal wave energy evolves according to
 \begin{equation}
     \label{eq: energy cont}% Factor 2 because of factor 1/2 in def of V.
     \dot{E_\alpha}= \sum_{\beta,\gamma}V_\alpha^{\beta \gamma}\Re \left(Z^*_\alpha Z^*_\beta Z^*_\gamma\right).
 \end{equation} 
The kinetic equation describes the evolution of $n_\alpha=\mean{E_\alpha}$, where the overbar denotes averaging over a suitable statistical ensemble.  In particular, we average over random Gaussian initial conditions such that
\begin{equation}
  \label{eq:initial data}
%\quad \mean{\Zb(0)\Za(0)} = 0,\,\,\, %\quand
  \mean{\Zb^*(0)\Za(0)} = \delta_{\alpha\beta} n_\alpha(0)
\end{equation}
is the only nonzero correlation.   The standard assumptions and procedural steps of weak wave turbulence  \cite{ZLF,buckmaster} then result in 
\begin{align}
\label{eq:cont}
&\dot{n_\alpha}=  \pi\!\!\int\limits_{\omega_{\alpha\beta\gamma}}\!\!V_{\alpha}^{\beta\gamma}\left(V_{\beta}^{\alpha\gamma}n_{\alpha}n_{\gamma}+V_{\gamma}^{\alpha\beta}n_{\beta}n_{\alpha}+V_{\alpha}^{\beta\gamma}n_{\beta}n_{\gamma}\right).
\end{align}
Here the joint kinetic limits of big box and long nonlinear times, $L\rightarrow\infty$ and $t\omega\rightarrow\infty$, were taken. So the discrete sums in \eqref{eq:z} and \eqref{eq: energy cont} were replaced by an integral over the resonant manifold %\addOB{I've added the wavenumber delta function here.  Previously it was incorporated into the definition of the interaction coeffcients, but those enter the kinetic equation in squared form, so that does not make sense.}
\begin{equation}
    \int\limits_{\omega_{\alpha\beta\gamma}} \!\!\!\!\!= \!\int d\beta d\gamma \,\delta(\omega_\alpha+\omega_\beta+\omega_\gamma)\delta\left(\mathbf{k}_{\alpha}+\mathbf{k}_{\beta}+\mathbf{k}_{\gamma}\right), 
    \end{equation}
where $\int\!\! d\alpha=\sum_{\sigma=\pm 1} \int\!\!d\mathbf{k}$.  The kinetic equation (\ref{eq:cont}) is generic for three-wave interactions, but it can be greatly simplified because of the non-generic additional conservation law for $P$.
The conservation of $E$ and $P$ for every triad (even non-resonant triads) implies
\begin{align}
  \label{eq:constraints}%and eq:9
  V_\alpha^{\beta\gamma}+V_\beta^{\alpha\gamma} +V_\gamma^{\beta\alpha} &= 0 \\
   s_\alpha V_\alpha^{\beta\gamma}+s_\beta  V_\beta^{\alpha\gamma} +s_\gamma V_\gamma^{\beta\alpha}&=0,\nonumber
\end{align}
respectively \cite{hasselmann1966feynman,Vann2005}.  
Notably, \eqref{eq:constraints} ensures conservation of $E$ and $P$ for any projection of \eqref{eq:z} onto a truncated set of modes. %The relations (\ref{eq:7},\ref{eq:8}) imply that truncated systems of three waves are integrable and equivalent to Euler's rigid body motion.
Viewing \eqref{eq:constraints} as dot products in
$\mathbb{R}^3$ means that $\vec{V}=(V_\alpha^{\beta\gamma},V_\beta^{\alpha\gamma},V_\gamma^{\beta\alpha})$ is
orthogonal to both $(1,1,1)$ and $(s_\alpha,s_\beta,s_\gamma)$, which already determines the direction of $\vec{V}$ uniquely. This makes clear that any other exact quadratic invariant that is diagonalized by the linear eigenbasis $\left\{ e_\alpha\right\}$ must be a linear combination of $E$ and $P$.  In particular, this implies that the vertical pseudomomentum based on 
the other slowness component cannot be exactly conserved by the full dynamics. 
However, the kinetic equation is restricted to the resonant
manifold $\omega_\alpha+\omega_\beta+\omega_\gamma =0$, %$\omega_{\alpha \beta \gamma }=0$, 
and therefore $\vec{\omega}=(\oma,\omb,\omg)$ is also orthogonal to both
$(1,1,1)$ and $(s_\alpha,s_\beta,s_\gamma)$, the latter because $s_\alpha\oma + s_\beta\omb+s_\gamma\omg =
 \hat{\bx}\cdot(\mathbf{k}_\alpha+\mathbf{k}_\beta+\mathbf{k}_\gamma)= 0.$ This means that $\vec{V}$ and $\vec{\omega}$ are
parallel to each other, i.e.,
\begin{equation}
  \label{eq:23}
  (V_\alpha^{\beta\gamma}, V_\beta^{\alpha\gamma}, V_\gamma^{\beta\alpha}) = \Gamma_{\alpha\beta\gamma} (\oma,\omb,\omg)
\end{equation}
for some real $ \Gamma_{\alpha\beta\gamma}$ totally symmetric in its indices.  For our system %\addOB{The defintion of the gammas can t have delta functions in it, not least because it enters the kinetic equation in squared form.  So I removed it.  I am unsure whether another factor of two is needed in its derivation.}
\begin{align}\label{eq:gamma}
\Gamma_{\alpha\beta\gamma}=\frac{\left(\sin\theta_{\alpha}+\sin\theta_{\beta}+\sin\theta_{\gamma}\right)}{\sqrt{8}}\left(\sigma_{\alpha}K_{\alpha}+\sigma_{\beta}K_{\beta}+\sigma_{\gamma}K_{\gamma}\right).%\delta\left(\mathbf{k}_{\gamma}+\mathbf{k}_{\alpha}+\mathbf{k}_{\beta}\right)
\end{align}
This changes \eqref{eq:cont} to 
\begin{equation}
      \label{eq:kin sim}
  \dot \na={\pi}\!\!
\int\limits_{\omega_{\alpha\beta\gamma}}\!\!\oma\,\Gamma_{\alpha\beta\gamma}^2 (\oma\nb\nc + \omb\na\nc +
    \omg\na\nb).%\delta(\omega_{\alpha \beta\gamma })
\end{equation}
Compared to \eqref{eq:cont}, this is a huge simplification and resembles the structure of kinetic equations derived for canonical Hamiltonian systems (including a similar entropy function $H(t)=\int d\alpha \log n_\alpha$). %(this also motivates considering an entropy function like $H(t)=\int d\alpha \log n_\alpha$). 
It is now apparent that on the resonant manifold additional conservation laws hold compared to the full system: any component of pseudomomentum %\addMS{(it will be interesting to understand if this component can be expressed in terms of $\phi$)} 
is now conserved, %if the corresponding wavenumbers add up to zero, 
so in $\mathbb{R}^d$ there are $d-1$ new
conservation laws that are valid for the kinetic equation. In particular, for the two-dimensional Boussinesq system the vertical pseudomomentum $P^z=\int d\alpha s^z_\alpha n_\alpha$ is conserved by the kinetic equation but not by the full flow; here $s^z_\alpha=k_z/\omega_\alpha$. 

Let us comment on the validity of the kinetic equation with respect to our initial assumption \eqref{eq:initial data}. The off-diagonal correlator $\overline{Z_{(+,\mathbf{k})}Z_{(-,\mathbf{k})}^*}$
%$\overline{Z_{\sigma_i}\left(\mathbf{k}\right)Z^*_{\sigma_j}\left(\mathbf{k}\right)}$ where $\sigma_i$ and $\sigma_j$ denote two different branches when $i\neq j$ 
has $O(\epsilon^2)$ fluctuations with frequency $2\omega_{+}(k)$. As long as this beating frequency does not vanish, the anomalous correlator \cite{zaleski2020anomalous} averages over time weakly to zero, so that the kinetic equation 
remains valid. The same holds for the correlator associated with space homogeneity $\overline{Z_{(\sigma,\mathbf{k})}^2}$. This is discussed in detail in the supplement together with a detailed derivation of the kinetic equation. 

\textbf{\textit{Steady solutions.}} 
The frequency \eqref{eq:omega} and the  coefficients \eqref{eq:23} are homogeneous functions of the wavenumbers  of degree $w_\omega=0$ and  $w_V=1$, respectively. That is, for $\lambda>0$, 
\begin{align}
\omega_{\left(\sigma,\lambda\mathbf{k}\right)}&=\omega_{\left(\sigma,\mathbf{k}\right)} \\
%\Gamma\left(\lambda\mathbf{k}_{\alpha},\lambda\mathbf{k}_{\beta},\lambda\mathbf{k}_{\gamma}\right)&=\lambda\Gamma\left(\mathbf{k}_{\alpha},\mathbf{k}_{\beta},\mathbf{k}_{\gamma}\right).
V_{\lambda\mathbf{k}_{\alpha}}^{\lambda\mathbf{k}_{\beta}\lambda\mathbf{k}_{\gamma}}&=\lambda V_{\mathbf{k}_{\alpha}}^{\mathbf{k}_{\beta}\mathbf{k}_{\gamma}}.
\end{align}
Formally at least, we can look for homogeneous solutions to the steady kinetic equation \eqref{eq:kin sim}, for which the steady spectrum is then of the separable form 
\begin{align}\label{eq:spectrum}
n_{\alpha}=n_\alpha^r(K)n_{\alpha}^{\Omega}\left(\theta_{k}\right). 
\end{align} 
Here $n^r_\alpha=K^{-w}$ for some suitable $w$. 
% Note that the interaction coefficient and frequency are also separable functions in angles and wave number amplitudes,
% \begin{align}
% \omega_{\alpha}&=\omega_{\alpha}^{r}\left(K\right)\omega_{\alpha}^{\Omega}\left(\theta_{k}\right)\\\Gamma_{\alpha\beta\gamma}&=\Gamma_{\alpha\beta\gamma}^{r}\left(K_{\alpha},K_{\beta},K_{\gamma}\right)\Gamma_{\alpha\beta\gamma}^{\Omega}\left(\theta_{\alpha},\theta_{\beta},\theta_{\gamma}\right).
% \end{align}
A trivial such formal steady solution is equipartition of energy such that $w = 0$ and $n^\Omega_\alpha=$const.  
%which follows from energy conservation,(\ref{eq:8}). %as any summand in the collision integral of
%the right hand side of (\ref{eq:kin}), is zero.
%For particular cases where the slowness is bounded from above or below, the kinetic equation has the solution $n_{\alpha}=\frac{1}{1+T_x^{-1}s^x_{\alpha}+T_z^{-1}s^z_{\alpha}+...}$, which %is isotropic, as $s_\alpha=\sigma N^{-1} K$,  and corresponds to equipartition of energy and pseudomomentum. These solutions are isotropic if the slowness is isotropic, as in the case of Rossby and internal gravity waves. 
This well-known solution has zero spectral flux of wave energy.  Non-zero flux steady solutions are the turbulent solutions, which are in general not isotropic.  However, regardless of their angular nature, as long as the forcing and dissipation are located at very distinct scales, an unambiguous spectral energy flux can be defined by integrating \eqref{eq:kin sim} over spheres of radius $K$. 
% \begin{equation}\label{eq:full continuity radial}
% \dot{n_\alpha^r}+K^{1-d}\partial_K(K^{d-1}\Pi_{K_{\alpha}}^r)=0,
% 	%\dot{n_\alpha^r}+K^{1-d}\partial_k(K^{d-1}\Pi_{k_{\alpha}}^r)=0,
%\end{equation} 
%Here $\Pi_{K_\alpha}^r=\int d\theta_k\Pi\cdot\hat{K}$  is the isotropic radial component of the flux.%, the product $K^{d-1}\Pi_{K_{\alpha}}^r$ is a non-zero constant in steady turbulence.
 
 Our result here is that
 \begin{equation}\label{eq:KZ spectra internal}
     n_\alpha^r=K^{-3}
 \end{equation} 
 is a steady solution that gives a constant nonzero  flux in two distinct cases.  First, $n_{(-,\mathbf{k})}=n_{(+,\mathbf{k})}$, so $P=0$ on average and there is symmetry between left-going and right-going waves.  Second, in cases where the pseudomomentum is concentrated on just one of the branches, e.g $n_{(-,\mathbf{k})}=0$, we actually find the two possible power laws
\begin{equation}\label{eq:KZ spectra internal PM}
n^{r}_E(K)=K^{-3},\,\,\, n^{r}_{PM}(K)=K^{-3.5},
\end{equation}
which correspond to energy and pseudomomentum cascades, respectively.  In both cases the isotropic spectrum of two-dimensional internal gravity waves is of the finite capacity type \cite{connaughton2010dynamical, krstulovic2023initial}, which shapes the time-dependent self-similar formation of the steady spectrum and might be of relevance to finite time singularity formation. See the supplement for a detailed derivation of these results.
%For particular cases where the slowness is bounded from above or below, the kinetic equation has the solution $n_{\alpha}=\frac{1}{1+T_x^{-1}s^x_{\alpha}+T_z^{-1}s^z_{\alpha}+...}$, which %is isotropic, as $s_\alpha=\sigma N^{-1} K$,  and corresponds to equipartition of energy andpseudomomentum. These solutions are isotropic if the slowness is isotropic, as in the case of Rossby and internal gravity waves. 

\textbf{\textit{Higher dimensions.}} 
We can generalize our results for systems of the form \eqref{eq:z} in $d$-dimensions, with two quadratic invariants: the energy, \eqref{eq:energy z}, and the pseudomomentum, \eqref{eq:PM z}. For $P=0$ symmetrically distributed between the frequency branches, $n_{(\sigma_i,\mathbf{k})}=n_{(\sigma_j,\mathbf{k})}$,  we obtain 
\begin{equation}\label{eq:KZ spectra}
n_E^{r}(K)=K^{-w^F_{E}},\,\,\,\, w^F_{E}=w_{V}+d-w_{\omega}/2.
\end{equation}
Here $w_V$ and $w_\omega$ are the homogeneity degrees of the interaction coefficients $V_\alpha^{\beta \gamma}$ and of the frequency $\omega_\alpha$.
%$If the interaction coefficient $\Gamma$ is homogeneous wrt $(K_\alpha,K_\beta,K_\gamma)$ of degree $w_\Gamma$ and the dispersion relation $\omega_\alpha$ is homogeneous 
%Note that $n_k$ is the energy density, rather than the number density, so 
Notably, the isotropic part of the energy spectrum of an anisotropic system gets an additional contribution of $w_\omega/2$ with respect to the Kolmogorov-Zakharov power law, $w_{KZ}=d+w_V-w_\omega$, of Hamiltonian isotropic three-wave interaction systems  \cite{ZLF,zak}.
Conversely, if the pseudomomentum is concentrated on one of the branches  %and the angular part of the frequency, $\omega^\Omega_\alpha$, can be viewed as a point on the unit circle, 
we find the additional scaling 
\begin{equation}\label{eq:PM spectrum}
    n_{PM}^{r}(K)=K^{-w^F_{PM}},\,\,\,\, w^F_{PM}=w_{V}+d+(1-2 w_\omega)/2,
\end{equation} 
which corresponds to a cascade of pseudomomentum. %so that \eqref{eq:KZ spectra internal PM} is a particular case for the 2D Boussinesq. 

\textbf{\textit{Relevance to other systems.}} Our results apply to systems  described by an equation of the general form \eqref{eq:e general eqn} or equivalently \eqref{eq:z}. Rossby waves in the mid latitude beta plane, on scales smaller compared to the deformation Radius, are governed by 
\begin{equation}\label{eq:rossby}
    \Delta\psi_{t}+\left\{ \psi,\Delta\psi\right\} =\beta\partial_{x}\psi 
\end{equation}
where $\psi$ is the stream function on the plane, so $-\Delta \psi$ is the vorticity. $x$ and $z$ are the zonal and meridional position coordinates and $f=f_0+\beta z$ is the Coriolis parameter. \eqref{eq:rossby} can be written in the form \eqref{eq:e general eqn} with $\phi=\psi, D=-\Delta, \mathcal{J}=\left\{ -\Delta\psi,\cdot\right\}, \mathcal{L}=\beta \partial_x$. It conserves the energy \eqref{eq:energy} and the pseudomomentum, \eqref{eq:pm} with $C=1$, which is the sign-definite enstrophy. Rossby drift waves in plasma are described by a similar equation. The wave expansion \eqref{eq:lin eigen} contains only one branch, $\alpha=\mathbf{k}$, so it is simply the Fourier transform. While the homogeneity degree of the interaction is the same as for internal gravity waves, the dispersion relation $\omega_k=-\beta\cos\theta/K$ is homogeneous of degree $w_\omega=-1$. Then
\eqref{eq:KZ spectra} and \eqref{eq:PM spectrum}  give the spectrum
%\begin{equation}\label{eq:rossby spectra}
 %   n_k^r\propto K^{-3.5},\,\,\,n_k^r\propto K^{-4.5}
%\end{equation} 
\begin{equation}\label{eq:rossby spectra}
    n_k^r\propto K^{-3.5},\,\,\,n_k^r\propto K^{-4.5}
\end{equation} 
for energy and enstrophy cascades, respectively. These scalings agree with the isotropic part of the unsteady spectra obtained by previous works \cite{pelinovsky1978wave}. Our work suggests to consider only the isotropic part of these solutions, while the angular part $n^\Omega_k$ can be obtained from the kinetic equation \eqref{eq:kin sim} after substituting the radial part $n^r_k$ from \eqref{eq:rossby spectra}. 

The dynamics of two-dimensional inertial waves in a vertical plane with constant Coriolis parameter $f$ is in fact governed precisely by the system we've studied in  \eqref{eq:e general eqn} after the replacement $x\longleftrightarrow z, f\longleftrightarrow N, \eta\longleftrightarrow v_2$. $\psi$ is the stream function on the vertical plane, $v_2$ is the velocity component perpendicular to the plane. Interestingly,  the vertical component of  pseudomomentum is then exactly conserved and is equal to the helicity of the flow.   The isotropic components of energy spectra are given by  \eqref{eq:KZ spectra internal} and \eqref{eq:KZ spectra internal PM} for the limiting cases of zero and sign-definite helicity, respectively. 

We are certainly not the first to apply weak wave turbulence approach to study internal gravity waves \cite{hasselmann1966feynman,mccomas1977resonant,muller1986nonlinear,caillol2000kinetic,lvov2001hamiltonian} and Rossby waves \cite{monin1987kinetic, connaughton2015rossby,pelinovsky1978wave, monin1987kinetic}. We believe that our work is the first example of theoretical prediction for internal gravity waves in $2D$ made by wave turbulence theory and has experimental practical relevance. Previous studies considered a narrow spectral range where the homogeneous wave number component is small compared to the non-homogeneous component. These yield spectral laws with diverging collision integrals, the divergence of the flux \eqref{eq:kin sim}, in the case of 3D internal gravity waves and are not stable in the case of Rossby waves \cite{balk1990physical} and hence cannot be physically realized. Our  solutions may represent the isotropic part of a physically relevant solution, while locality (convergence of the collision integrals) is ensured by obstructions on the complementary angular part of the energy spectrum $n_k^\Omega$ \eqref{eq:spectrum}. 

%The scaling $K^{-4}$ was numerically observed by \cite{danilov2004scaling}.

\textit{\textbf{Speculations on inverse energy cascade of unidirectional internal waves}}. The decomposition \eqref{eq:lin eigen} of the field $\phi=\sum_{\alpha}Z_{\alpha} e_{\alpha}$ into right-going and left-going waves splits the pseudomomentum \eqref{eq:PM z} into positive and negative components $P =P_{+}+P_{-}$, recall
that the positive horizontal slowness is $s^x_{+}=k_{x}/\omega_{+}=K/N$.  In the case of waves solely propagating to the right initially, $Z_{(-,\mathbf{k})}\left(t=0\right)=0$ for all $\mathbf{k}$, the pseudomomentum is positive at $t=0$, and its time derivative $\dot{P}_{+}\left(t=0\right)=0$ vanishes as well, as evident from \eqref{eq: energy cont}. The persistence of predominantly right-going  waves is fixed into the memory of the system and cannot be forgotten since the pseudomomentum is an exact invariant; so any generation of leftward propagating waves through nonlinear interactions must be accompanied by an equal creation of rightward propagating waves. This has practical relevance for ocean dynamics.  For example,  strongly directional internal wave fields arise naturally in the case  of internal tides radiated away from isolated topography structures such as the Hawaiian ridge \cite{LSY03}. During periods when unidirectional waves dominate, the system can be effectively described by a kinetic equation truncated to the positive branch, which shares a similar structure with the kinetic equation of Rossby waves. While it is well known that the energy of Rossby waves is transferred, albeit not isotropically, from small to large scales; an inverse energy cascade of internal gravity waves has never been observed. Ripa \cite{ripa} briefly mentions the idea of an inverse energy cascade of internal gravity waves, adapting the classic dual cascade argument \cite{fjortoft53}.   Similarly to two-dimensional hydrodynamics \cite{kraichnan1967inertial}, the mechanism that drives the inverse energy cascade of Rossby waves is the existence of a second sign definite quadratic invariant, the enstrophy, with a density proportional by an isotropic monotonic function. When unidirectional internal waves dominate, %and pseudomomentum becomes approximately sign-definite its density is proportional to that of the energy by the (now sign definite) slowness. In this scenario, 
the pseudomomentum can play a role akin to enstrophy by driving the energy up scale in the presence of a small or intermediate scale forcing. In this limit as $P_+/ P_-\rightarrow\infty$, we obtain the scalings for the radial energy spectra given by \eqref{eq:KZ spectra internal} and \eqref{eq:KZ spectra internal PM} for energy and pseudomomentum cascades, respectively. For isotropic systems it is widely accepted that the ordering of the scalings, $3=W^F_E<W^F_{PM}=3.5$, implies an inverse cascade of energy \cite{nazabook}. While energy transfer in non-isotropic systems can be intricate, this strongly suggest the existence of inverse energy cascade of internal waves in the effective description. The extent and time scales of this approximation for the 2D Boussinesq equation remain a topic for future study. 

\textbf{\textit{Adding rotation}}. The first step towards including both rotation and stratification in the kinetic equation for internal gravity waves \eqref{eq:kin sim} can be taken whilst retaining the two-dimensional nature of the flow, i.e., $\partial_y=0$.  This involves adding horizontal Coriolis forces to the momentum equations, which necessitates allowing for a third velocity component $v_2$ in the $y-$direction.  The Coriolis forces based on a constant Coriolis parameter $f$ add only linear terms to the governing equations, so the linear part of the dynamics is described by the 3x3 operator 
\begin{equation}%\nonumber
%\phi=\begin{pmatrix}A\\
%v_{2}\\
%\eta
%\end{pmatrix},\,\,\,
N^{2}\begin{pmatrix}0 & 0 & 1\\
0 & 0 & 0\\
1 & 0 & 0
\end{pmatrix}\partial_{x}+f\begin{pmatrix}0 & 1 & 0\\
1 & 0 & 0\\
0 & 0 & 0
\end{pmatrix}\partial_{z}.
\end{equation}
The state vector is now $\phi^T=(\psi,v_2,\eta)$ and the  Hermitian diagonal operator is 
$D=\text{diag}(-\Delta,1,N^2)$.
This generalizes \eqref{eq:2D Boussinesq} %\cite{ripa} 
to include the Coriolis force within the $f$-plane approximation, $f\hat{z}\cross\mathbf{v}$. The energy is $E=\braket{\phi}{D\phi}=\int d\mathbf{x}\left(-\psi \Delta \psi +v_2^2+N^2\zeta^{2}\right)$.  The dispersion relation is
\begin{equation}
\omega_{(\sigma,\mathbf{k})}=\sigma N\cos\theta_{k}\sqrt{1+f^{2}N^{-2}\tan^{2}\theta_{k}}    
\end{equation}
with $\sigma=0,\pm$. The expansion of $\phi$ in terms of eastward and westward propagating waves, \eqref{eq:lin eigen}, leads to the kinetic equation \eqref{eq:cont} with the interaction coefficients %\addOB{Again should probably not include the delta function}
\begin{align}
V_\alpha^{\beta\gamma}&=-\frac{\mathbf{k}_{\beta}\cross\mathbf{k}_{\gamma}}{2\sqrt{8}K_{\alpha}K_{\beta}K_{\gamma}} \cdot
\\
&\left(K_{\beta}^{2}-K_{\gamma}^{2}+f^{2}s_{\alpha}^{z}\left(s_{\gamma}^{z}-s_{\beta}^{z}\right)+N^{2}s_{\alpha}^{x}\left(s_{\gamma}^{x}-s_{\beta}^{x}\right)\right).\nonumber
\end{align}
%A small uniform parameter relevant for internal wave interaction is the root mean square vertical gradient of the vertical displacement $\epsilon  =\sqrt{\left\langle \left(\partial_{z}\zeta\right)^{2}\right\rangle }.$ 
This does not include interactions among and with the zero frequency branch, also known as the balanced or shear modes. Formally, at the limit of vanishing frequency, off-diagonal correlators should be added as well to the kinetic equation. However, weak wave turbulence closure is not expected to remain valid when shear vortical modes carry the dominant part of the energy \cite{labarre2023internal}. Energy is an exact invariant, so the first constraint in \eqref{eq:constraints} holds. As the homogeneity degree of the dispersion relation remains $w_\omega=0$, our finding \eqref{eq:KZ spectra}, suggests that rotation only modifies the angular component of the turbulent energy spectrum \eqref{eq:spectrum}, but leaves the radial component, \eqref{eq:KZ spectra internal}, unchanged. 
%When stratification dominates over rotation, quantified by a small Rossby number $R_0=v/Lf$, the horizontal pseudomomentum is an exact invariant $P^x=\int d\alpha \sigma N^{-1}K  Z_\alpha Z_\alpha^*$. When rotation dominates over stratification, the vertical pseudomomentum is an exact invariant $P^z=\int d\alpha \sigma f^{-1}K Z_\alpha Z_\alpha^*$. In both of these limits the kinetic equation is simplified to the form (\ref{eq:kin sim}) and conserves both Casimirs - i.e the two components of the pseudomomentum. 
We note that the conservation of potential vorticity might be used to simplify the kinetic equation in the presence of both rotation and stratification. This is studied in a future work.
 
% (\addMS{MS: should we discuss here the relation to the Garrett Munk spectrum and physical measurements}).

% Note that equations (\ref{eq:z},\ref{eq:energy co},\ref{eq:PM co})
%describe the Boussinesq system in the form of a hydrodynamic type system
%with two integrals of motion \cite{Obukhov,hydro two}. 

\textit{\textbf{Conclusion.}}
%It is clear that wave interactions and the sterile approach of wave turbulence theory cannot fully capture the intricate dynamics of geophysical systems nor precisely predict the energy spectra measured in the ocean. %and the atmosphere. 
%Our work is motivated by both practical and theoretical considerations. 
Our work emphasizes the elegant ramifications in the theoretical description and in practice of sign-indefinite invariants, which usually do not get much attention in wave turbulence study. We show that the existence of a second quadratic invariant, simplifies the kinetic equation and leads to additional conservation laws on the resonant manifold, which to our knowledge, were previously unknown in the geophysical community. This simplification facilitates the derivation of scaling laws for the isotropic component of the turbulent wave spectra of 2D internal gravity waves and Rossby waves. We show that there are practical scenarios in which pseudomomentum conservation can drive an inverse energy cascade of internal gravity waves. On the theoretical front, our work contributes a different approach to the study of wave turbulence in non-isotropic systems dominated by three-wave interactions. This encompasses the application of non-canonical variables for deriving the kinetic equation, and using variable separation in order to find turbulent solutions of the kinetic equation.

\textit{\textbf{Acknowledgments}}.
%\begin{acknowledgments}
We thank Gregory Falkovich and Vincent Labarre for useful discussions. 
This work was supported by the Simons Foundation and the Simons Collaboration on Wave Turbulence.
  %The numerical study was made
  %possible thanks to New York University's Greene computing
  %cluster facility.
  OB acknowledges additional financial support under ONR grant N00014-19-1-2407 and NSF grant DMS-2108225. MS acknowledges additional financial support from the Schmidt Futures Foundation. %and the Israeli CHE. 
%\end{acknowledgments}

\bibliography{main}% Produces the bibliography via BibTeX.

\end{document}

\global\long\def\dyad#1{\underline{\underline{\boldsymbol{#1}}}}%
\global\long\def\ubar#1{\underbar{\ensuremath{\boldsymbol{#1}}}}%
\global\long\def\integer{\mathbb{Z}}%
\global\long\def\natural{\mathbb{N}}%
\global\long\def\real#1{\mathbb{R}^{#1}}%
\global\long\def\complex#1{\mathbb{C}^{#1}}%
\global\long\def\defined{\triangleq}%
\global\long\def\trace{\text{trace}}%
\global\long\def\del{\nabla}%
\global\long\def\cross{\times}%
\global\long\def\diff#1#2{\frac{\partial#1}{\partial#2}}%
\global\long\def\Diff#1#2{\frac{d#1}{d#2}}%
\global\long\def\bra#1{\left\langle #1\right|}%
\global\long\def\ket#1{\left|#1\right\rangle }%
\global\long\def\braket#1#2{\left\langle #1|#2\right\rangle }%
\global\long\def\ketbra#1#2{\left|#1\right\rangle \left\langle #2\right|}%
\global\long\def\identity{\mathbf{1}}%
\global\long\def\paulix{\begin{pmatrix}  &  1\\
 1 
\end{pmatrix}}%
\global\long\def\pauliy{\begin{pmatrix}  &  -i\\
 i 
\end{pmatrix}}%
\global\long\def\pauliz{\begin{pmatrix}1\\
  &  -1 
\end{pmatrix}}%
\global\long\def\sinc{\mbox{sinc}}%
\global\long\def\four{\mathcal{F}}%
\global\long\def\dag{^{\dagger}}%
\global\long\def\norm#1{\left\Vert #1\right\Vert }%
\global\long\def\hamil{\mathcal{H}}%
\global\long\def\tens{\otimes}%
\global\long\def\ord#1{\mathcal{O}\left(#1\right)}%
\global\long\def\undercom#1#2{\underset{_{#2}}{\underbrace{#1}}}%
 
\global\long\def\conv#1#2{\underset{_{#1\rightarrow#2}}{\longrightarrow}}%
\global\long\def\tg{^{\prime}}%
\global\long\def\ttg{^{\prime\prime}}%
\global\long\def\clop#1{\left[#1\right)}%
\global\long\def\opcl#1{\left(#1\right]}%
\global\long\def\broket#1#2#3{\bra{#1}#2\ket{#3}}%
\global\long\def\div{\del\cdot}%
\global\long\def\rot{\del\cross}%
\global\long\def\up{\uparrow}%
\global\long\def\down{\downarrow}%
\global\long\def\Tr{\mbox{Tr}}%

\global\long\def\per{\mbox{}}%
\global\long\def\pd{\mbox{}}%
\global\long\def\p{\mbox{}}%
\global\long\def\ad{\mbox{}}%
\global\long\def\a{\mbox{}}%
\global\long\def\la{\mbox{\ensuremath{\mathcal{L}}}}%
\global\long\def\cm{\mathcal{M}}%
\global\long\def\cg{\mbox{\ensuremath{\mathcal{G}}}}%

\title{Supplemental Material for The role of sign indefinite invariants in shaping turbulent cascades}
\author{Michal Shavit, Oliver B\"uhler 
and Jalal Shatah\\
\emph{\normalsize{}Courant Institute of Mathematical Sciences, New
York University, NY 10012, USA }}
\maketitle

\tableofcontents{}

\section{Closure leading to  kinetic equation}
\label{sec:clos-kinet-equat}
Let us rewrite the wave evolution equation, (10) in the main text, in terms of small-amplitude waves,
\begin{equation}\label{eq:z epsilon}
  \dot \Za+i\oma \Za = \eps \sum_{\beta,\gamma} \frac{1}{2}V_\alpha^{\beta \gamma}
  \Zb^*\Zg^*,
\end{equation}
The explicit small-amplitude  parameter $0<\eps\ll1$
makes $\Za=O(1)$. Such small uniform parameter relevant for internal wave interaction is the root mean square vertical gradient of the vertical displacement $\epsilon  =\sqrt{\left\langle \left(\partial_{z}\zeta\right)^{2}\right\rangle }$.
The discussion can also be generalized to a non-uniform small parameter $\epsilon=\epsilon_{\alpha}$. %From () the kinetic equation can be derived without much effort. 
Assuming initial conditions are random, we are interested in writing an evolution
equation up to order $\epsilon^{2}$ to the averaged energy density
$n_{\alpha}:=\overline{(E_\alpha)} =\overline{(Z_{\alpha}Z_{\alpha}^{*})}$,
where $\overline{(\ldots)}$ denotes the average with respect to the initial data distribution.  The resulting kinetic equation tracks the statistical evolution of \eqref{eq:z epsilon} over very long times of $O(\eps^{-2})$. In order to derive the kinetic equation, we expand
the energy density $E_\alpha=Z_{\alpha}Z_{\alpha}^*$ in terms of the initial data using integration by parts. Transforming to envelopes $Z_{\alpha}\rightarrow Z_{\alpha}e^{-i\omega_{\alpha}t}$ the equation for the energy density is 
\begin{equation}\label{eq:energy density}
 \dot E_{\alpha}\left(t\right)=\epsilon\sum_{\beta,\gamma}V_{\alpha}^{\beta\gamma}Z_{\alpha}^{*}Z_{\beta}^{*}Z_{\gamma}^{*}e^{i\omega_{\alpha\beta\gamma}t}+c.c.
\end{equation}
%\begin{equation}\label{eq:third moment}
 %E_{\alpha}\left(t\right)=E_{\alpha}\left(0\right)+\epsilon\int_{0}^{t}ds\sum_{\beta,\gamma}\sigma_{\alpha}^{\beta\gamma}Z_{\alpha}^{*}Z_{\beta}^{*}Z_{\gamma}^{*}e^{i\omega_{\alpha\beta\gamma}t}+c.c.
%\end{equation}
Laplace transform and integrate by parts once 
\begin{align}\label{eq:Lap1}
    s\tilde{E}_{\alpha}\left(s\right)-E_{\alpha}\left(0\right)&=\epsilon\int_{0}^{\infty}\sum_{\beta,\gamma}V_{\alpha}^{\beta\gamma}Z_{\alpha}^{*}Z_{\beta}^{*}Z_{\gamma}^{*}e^{-\left(s-i\omega_{\alpha\beta\gamma}\right)t}dt+c.c,\nonumber
    \\
    &=\epsilon P_{3}\mid_{t=0}+\epsilon\int_{0}^{\infty}\sum_{\beta,\gamma}V_{\alpha}^{\beta\gamma}
    \frac{d}{dt}(Z_{\alpha}^{*}Z_{\beta}^{*}Z_{\gamma}^{*})
    \frac{e^{-(s-i\omega_{\alpha\beta\gamma})t}}{s-i\omega_{\alpha \beta \gamma}}dt+c.c,
\end{align}
where $\tilde{E}(s)$ denotes the Laplace transform of $E_\alpha(t)$ and $P_3$ stands for polynomials of third order in the amplitudes. Further, we write the time derivative of the third moment
\begin{equation}\label{eq:3rd moment}
    \frac{d}{dt}\left(Z_{\alpha}^{*}Z_{\beta}^{*}Z_{\gamma}^{*}\right)=\epsilon\sum_{\lambda\kappa}\frac{1}{2}V_{\beta}^{\lambda\kappa}\left(Z_{\lambda}Z_{\kappa}Z_{\gamma}^{*}Z_{\alpha}^{*}\right)e^{-i\omega_{\kappa\lambda\beta}t}+\text{permutations}\left\{ \alpha\rightarrow\gamma,\alpha\rightarrow\beta\right\} .
\end{equation}
\eqref{eq:energy density} and \eqref{eq:3rd moment} are exact. We integrate by parts \eqref{eq:Lap1} once more, substitute \eqref{eq:3rd moment} and arrive at
\begin{equation}\label{eq:Lap2}
s\tilde{E}_{\alpha}\left(s\right)-E_{\alpha}\left(0\right)=P_{3}\mid_{t=0}+\epsilon^{2}\sum_{\beta,\gamma,\lambda\kappa}\frac{1}{2}V_{\alpha}^{\beta\gamma}V_{\beta}^{\lambda\kappa}\frac{\left(Z_{\lambda}Z_{\kappa}Z_{\gamma}^{*}Z_{\alpha}^{*}\right)\mid_{t=0}}{\left(i\omega_{\alpha\beta\gamma}-s\right)\left(i\omega_{\alpha\beta\gamma}-i\omega_{\kappa\lambda\beta}-s\right)}+\text{permutations}+c.c.+O\left(\epsilon^{3}\right)
\end{equation}
We assume the initial distribution of amplitudes is exactly Gaussian, 
\begin{equation}
  \label{eq:initial data}
  \mean{\Zb^*(0)\Za(0)} = \delta_{\alpha\beta} n_\alpha(0),
\end{equation}
and the rest of the correlators are zero. 
\begin{comment}
\begin{equation}
  \label{eq:initial data}
  \mean{\Za(0)} = 0, \quad \mean{\Zb(0)\Za(0)} = 0, \quand
  \mean{\Zb^*(0)\Za(0)} = \delta_{\alpha\beta} n_\alpha(0). 
\end{equation}
\end{comment}
We now take the inverse Laplace transform of \eqref{eq:Lap2}, average with respect to the initial conditions \eqref{eq:initial data} and at the equation for the average energy density
\begin{align}\label{eq:pre kin}
n_{\alpha}(t) =n_{\alpha}(0) +\epsilon^{2}\sum_{\alpha,\beta}\frac{1}{2}V_{\alpha}^{\beta\gamma}\left(V_{\beta}^{\alpha\gamma}n_{\alpha}n_{\gamma}+V_{\gamma}^{\alpha\beta}n_{\beta}n_{\alpha}+V_{\alpha}^{\beta\gamma}n_{\beta}n_{\gamma}\right)\frac{1-\cos\omega_{\alpha\beta\gamma}t}{\omega_{\alpha\beta\gamma}^{2}}+O\left(\epsilon^{4}\right).
\end{align}
The expansion includes only even powers of $\epsilon$, since odd powers correspond
to odd monomials of the amplitudes $Z_{\alpha}$, which vanish due to the Gaussianity of the initial distribution. %In order to derive the kinetic equation one should start from the discrete case of a finite domain $\Omega=\left[0,L\right]$ and 
We take the kinetic limits of big box $L\rightarrow\infty$ and and long times $t\omega\rightarrow\infty$ assuming that there are enough quasi-resonances with respect to exact resonances so the derivation occurs.  The time fluctuating factor is replaced by $\lim_{t\rightarrow\infty}\frac{1-\cos\omega_{\alpha\beta\gamma}t}{\omega_{\alpha\beta\gamma}^{2}}=2\pi t\delta\left(\omega_{\alpha\beta\gamma}\right)$ in the sense of distributions and we arrive at the kinetic equation, (14) in the main text, 
\begin{align}\label{eq:kin}
\dot{n_{\alpha}} & =\pi\epsilon^{2}\int\!\!    V_{\alpha}^{\beta\gamma}\left(V_{\beta}^{\alpha\gamma}n_{\alpha}n_{\gamma}+V_{\gamma}^{\alpha\beta}n_{\beta}n_{\alpha}+V_{\alpha}^{\beta\gamma}n_{\beta}n_{\gamma}\right)\delta\left(\omega_{\alpha\beta\gamma}\right)d\beta d\gamma.
\end{align}
%In taking the limit $\lim_{t\omega\rightarrow\infty}\frac{1-\cos\omega_{\alpha\beta\gamma}t}{\omega_{\alpha\beta\gamma}^{2}}=\pi t\delta\left(\omega_{\alpha\beta\gamma}\right)$
%we assume that the number of quasiresonances is very large compared
%to exact resonances. 
The discrete sums over $\na(t)$ in \eqref{eq:pre kin} are replaced by integrals over a continuous measure
$n_\alpha(t)\,\mathrm{d}\alpha$, where $\int\!\! d\alpha=\int\!\! d\sigma \int\!\!d\mathbf{k}$ and  $ d \sigma=\sum_{\sigma_i}$ is the counting measure over the frequency branches.  Now integration over the
resonant manifold is defined by $\omega_{\alpha \beta\gamma}=\oma+\omb+\omg = 0$. 

\begin{comment}
The off diagonal correlator $\overline{Z_{\sigma_i}\left(\mathbf{k}\right)Z^*_{\sigma_j}\left(\mathbf{k}\right)}$ where $\sigma_i$ and $\sigma_j$ denote two different branches when $i\neq j$ has $O(\epsilon^2)$ fluctuations with frequency $\omega_{\sigma_i}(k)-\omega_{\sigma_j}(k)$. As long as this beating frequency does not vanish, the anomalous correlator averages over time weakly to zero, so that the kinetic equation (\ref{eq:kin})
remains valid. This is discussed in detail in the appendix. All other off diagonal correlators remain identically zero. 
\end{comment}

\section{Derivation of the isotropic turbulent spectra}
For the derivation of the isotropic turbulent spectra, (13) and (14) in the main text, we consider the following very easy to prove property: Let $f:\mathbb{R}^{+}\times\mathbb{R}^{+}\rightarrow\mathbb{R}^{+}$
be a homogeneous function of degree $0$, that is $\forall\lambda\in\mathbb{R}^{+}$
$f\left(\lambda k_{1},\lambda k_{2}\right)=f\left(k_{1},k_{2}\right)$.
If $\forall k_{1}\in\mathbb{R}^{+}$, $f\left(k_{1},k_{2}\right)k_{2}^{-1}\in L_{1}\left(\mathbb{R}^{+}\right)$
then 
\begin{align}
\frac{d}{dk_{1}}\int_{0}^{\infty}f\left(k_{1},k_{2}\right)\frac{dk_{2}}{k_{2}} & =0,
\end{align}
in particular, (AM Balk, 2000), 
\begin{align}\label{balk}
\int_{0}^{\infty}f\left(k_{1},k_{2}\right)\frac{dk_{2}}{k_{2}} & =\int_{0}^{\infty}f\left(k_{2},k_{1}\right)\frac{dk_{2}}{k_{2}}.
\end{align}
For simplicity, let us consider a kinetic equation with one branch
(e.g for eastward propagating internal waves or for Rossby waves) 

%\begin{align}
%\dot{n}_{k}+\text{div}\Pi_{k} & =0,\label{eq:app kin pos-1-2-1}\end{align}
\begin{align}
\dot{n}_{k} & =\epsilon^{2}\pi\int d\mathbf{q}\int d\mathbf{p}V_{k}^{pq}\left(V_{k}^{pq}n_{p}n_{q}+V_{p}^{kq}n_{k}n_{q}+V_{q}^{pk}n_{p}n_{k}\right)\delta\left(\omega_{p,q,k}\right)\delta\left(\mathbf{k}+\mathbf{q}+\mathbf{p}\right).\label{eq:app kin pos-1-2}
\end{align}
Assume that interaction coefficients and frequencies are homogeneous
functions of degree $w_{V}$ and $w_{\omega}$, respectively. That
is, for $\lambda>0$, 
\begin{align}
\omega\left(\lambda\mathbf{k}\right) & =\lambda^{w_{\omega}}\omega\left(\mathbf{k}\right)\\
%\Gamma\left(\lambda\mathbf{k},\lambda\mathbf{q},\lambda\mathbf{q}\right) & =\lambda^{w_{\Gamma}}\Gamma\left(\mathbf{k},\mathbf{q},\mathbf{p}\right).
%\\
V_{\lambda\mathbf{k}_{\alpha}}^{\lambda\mathbf{k}_{\beta}\lambda\mathbf{k}_{\gamma}}&=\lambda^{w_{V}}V_{\mathbf{k}_{\alpha}}^{\mathbf{k}_{\beta}\mathbf{k}_{\gamma}}
\end{align}
So, it makes sense to look for steady homogeneous solutions of the kinetic equation. The latter implies that the steady spectrum is a
separable function in wave number amplitude and solid angle
\begin{align}
n_{k} & =n_{k}^{r}\left(K\right)n_{k}^{\Omega}\left(\Omega_{k}\right),\label{eq:spectrum}
\end{align}
with $n_{k}^{r}=K^{-w}$. Here $\Omega_{k}$ denotes the $d-1$
unit sphere. We write the collision integral in the following form:
\begin{align}
\dot{n}_{k} =K^{y}\int\frac{dK_{q}}{K_{q}}\int\frac{dK_{p}}{K_{p}}\int d\Omega_{pq}V_{k}^{pq}K^{-y-d}U\left(K,\Omega_{k},K_{p},\Omega_{p},K_{q},\Omega_{q}\right),
\end{align}
where $d\Omega_{pq}=d\Omega_{p}d\Omega_{q}$ and 
\begin{equation}
U\left(K,\Omega_{k},K_{p},\Omega_{p},K_{q},\Omega_{q}\right)=\epsilon^{2}\pi K^{d}K_{p}^{d}K_{q}^{d}\left(V_{k}^{pq}n_{p}n_{q}+V_{p}^{kq}n_{k}n_{q}+V_{q}^{pk}n_{p}n_{k}\right)\delta\left(\omega_{p,q,k}\right)\delta\left(\mathbf{k}+\mathbf{q}+\mathbf{p}\right).
\end{equation}
We assume that $V_{k}^{pq}K^{-y-d}U$ is a homogeneous function (wrt
$K,K_{q},K_{p}$) of degree $0$. So $y$ amounts to a sum of the homogeneity degrees of the multiplicative factors of the collision integral. We write the collision integral as sum of three identical copies
\begin{align*}
\dot{n}_{k}=\frac{1}{3}K^{y}\int d\Omega_{pq}\int\frac{dK_{q}}{K_{q}}\int\frac{dK_{p}}{K_{p}} & (V_{k}^{pq}K^{-y-d}U\left(K,\Omega_{k},K_{p},\Omega_{p},K_{q},\Omega_{q}\right)\\
 & +V_{k}^{pq}K^{-y-d}U\left(K,\Omega_{k},K_{p},\Omega_{p},K_{q},\Omega_{q}\right)+V_{k}^{pq}K^{-y-d}U\left(K,\Omega_{k},K_{p},\Omega_{p},K_{q},\Omega_{q}\right)).
\end{align*}
Apply \eqref{balk} on the second line terms 
\begin{align*}
\dot{n}_{k}=\frac{1}{3}K^{y}\int d\Omega_{pq}\int\frac{dK_{q}}{K_{q}}\int\frac{dK_{p}}{K_{p}} & (V_{k}^{pq}K^{-y-d}U\left(K,\Omega_{k},K_{p},\Omega_{p},K_{q},\Omega_{q}\right)\\
 & +V_{p}^{kq}K_{p}^{-y-d}U\left(K_{p},\Omega_{k},K,\Omega_{p},K_{q},\Omega_{q}\right)+V_{q}^{pk}K_{q}^{-y-d}U\left(K_{q},\Omega_{k},K_{p},\Omega_{p},K,\Omega_{q}\right)).
\end{align*}
We now integrate the collision integral over the unit sphere $\int d\Omega_{k}$,
permute the angles in the second term $\Omega_{k}\longleftrightarrow\Omega_{p}$
and in the third term $\Omega_{k}\longleftrightarrow\Omega_{q}$ and
arrive at
\begin{align}
%-K^{1-d}\partial_{K}\left(K^{d-1}\Pi_{k}\right) & 
\dot n^r_k=\frac{1}{3}K^y\int d\Omega_{pq}\int\frac{dK_{q}}{K_{q}}\int\frac{dK_{p}}{K_{p}}\left[V_{k}^{pq}K^{-y-d}+V_{p}^{kq}K_{p}^{-y-d}+V_{q}^{pk}K_{q}^{-y-d}\right]U\left(K,\Omega_{k},K_{p},\Omega_{p},K_{q},\Omega_{q}\right).
\end{align}
When $y+d=0$ then the brackets are the energy conservation and hence
zero. This fixes the radial component of the spectrum
\begin{align}\label{E spectrum}
w_{E} & =d+w_{V}-\frac{1}{2}w_{\omega}.
\end{align}
so that $n_{k}^{r}=K^{-w_{E}}$ is the energy cascade formal solution
of the radial kinetic equation (the continuity equation integrated
over the angle).

Applying the constraint of pseudo-momentum conservation on the interaction
coefficient, $V_{k}^{pq}=\omega_{k}\Gamma_{pqk}$, the brackets become 
\begin{align*}
\dot n^r_k =\frac{1}{3}\int d\Omega_{pq}\int\frac{dK_{q}}{K_{q}}\int\frac{dK_{p}}{K_{p}}K^{y}\left[\omega_{k}K^{-y-d}+\omega_{p}K_{p}^{-y-d}+\omega_{q}K_{q}^{-y-d}\right]\Gamma_{pqk}U\left(K,\Omega_{k},K_{p},\Omega_{p},K_{q},\Omega_{q}\right).
\end{align*}
In cases when the angular component $\omega_{k}^{\Omega_{k}}\in\Omega_{k}$
is a point on the unit sphere, then if $w_{\omega}-y-d=1$ the brackets
are the space homogeneity condition and hence zero. This gives the radial component of the spectrum
\begin{align}\label{PM spectrum}
w_{PM} & =d+w_{V}-w_{\omega}+\frac{1}{2},
\end{align}
which is another formal solution of the radial kinetic equation in
cases when the pseudo-momentum is positive and corresponds to pseudo-momentum cascade. In terms of the homogeneity degree of the interaction coefficient $\Gamma$ which equals $w_{V}=w_{\Gamma}+w_{\omega}$, \eqref{E spectrum} and \eqref{PM spectrum} are given by 
\begin{align}
w_{E} & =d+w_{\Gamma}+\frac{1}{2}w_{\omega},\\
w_{PM} & =d+w_{\Gamma}+\frac{1}{2}.
\end{align}

%which are (13) and (14) in the main text.

\section{Off diagonal second order anomalous correlators}

In writing a kinetic theory for the averaged energy density $n_{\alpha}=\overline{Z_{\alpha}Z_{\alpha}^{*}}$
we assume random Gaussian initial conditions, (16) in the main text, where the only non-zero cumulants are
\begin{equation}
  \label{eq:initial data}
%\quad \mean{\Zb(0)\Za(0)} = 0,\,\,\, %\quand
  \mean{\Zb^*(0)\Za(0)} = \delta_{\alpha\beta} n_\alpha(0).
\end{equation}
One needs to make sure that the off diagonal correlators   $\overline{Z_{(+,\mathbf{k})}Z_{(-,\mathbf{k})}^*}$ and $\overline{Z_{(\sigma,\mathbf{k})}^2}$
remain zero, otherwise these correlators should be added to the kinetic equation
(\ref{eq:kin}) or at least monitored. Such off diagonal correlators sometimes referred to as anomalous correlators. %\cite{anom}. 
We carry on here the calculation for $\overline{Z_{(+,\mathbf{k})}Z_{(-,\mathbf{k})}^*}$ and show that it averages weakly to zero as long as $\omega_{\sigma,\mathbf{k}}>0$. In a simlar calculation, one can show the same for  correlator associated with space homogeneity $\overline{Z_{(\sigma,\mathbf{k})}^2}$.

Though initial conditions
are Gaussian the time derivative $\frac{d}{dt}\overline{Z_{+}Z_{-}^{*}} \neq0$
is not zero identically. We show that it is zero in the weak sense of
distributions; that is, rapidly fluctuating around zero in the kinetic
limit $t\omega\rightarrow\infty.$ Writing the correlator in terms of the Fourier expansion of the stream function and elevation:
\begin{align}
\overline{Z_{+}Z_{-}^{*}}&=\frac{1}{2}\overline{\left( K^2\hat{\psi}\left(\mathbf{k}\right)\hat{\psi}^{*}\left(\mathbf{k}\right)-N^{2}\hat{\zeta}\left(\mathbf{k}\right)\hat{\zeta}^{*}\left(\mathbf{k}\right)\right)}
+KN\overline{\left(\hat{\psi}\left(\mathbf{k}\right)\hat{\zeta}^{*}\left(\mathbf{k}\right)-\hat{\zeta}\left(\mathbf{k}\right)\hat{\psi}^{*}\left(\mathbf{k}\right)\right)},
\end{align}
we see that the physical interpretation of the case where the two brackets fluctuate each around zero in the equation above is that the kinetic energy stored at each wave number $\mathbf{k}$ equals to the potential energy stored at this wave number and that the pseudo-momentum stored at $\mathbf{k}$ equals to that stored at $\mathbf{-k}$.

We start by computing the time derivative of the product $Z_{\left(+,\mathbf{k}\right)}Z_{\left(-,\mathbf{k}\right)}^{*}$:
\begin{align}
\frac{d}{dt}\left(Z_{\left(+,\mathbf{k}\right)}Z_{\left(-,\mathbf{k}\right)}^{*}\right)/\epsilon & =\sum_{\beta,\gamma}V_{\left(+,\mathbf{k}\right)}^{\beta\gamma}Z_{\beta}^{*}Z_{\gamma}^{*}Z_{\left(-,\mathbf{k}\right)}^{*}e^{i\left(\omega_{\gamma}+\omega_{\beta}+\omega_{\left(+,\mathbf{k}\right)}\right)t}+\sum_{\beta,\gamma}V_{\left(-,\mathbf{k}\right)}^{\beta\gamma}Z_{\left(+,\mathbf{k}\right)}Z_{\beta}Z_{\gamma}e^{-i\left(\omega_{\gamma}+\omega_{\beta}+\omega_{\left(-,\mathbf{k}\right)}\right)t}.
\end{align}
To write the kinetic equation for $\overline{Z_{\left(+,\mathbf{k}\right)}Z_{\left(-,\mathbf{k}\right)}^{*}}$ we take the Laplace transform and integrate by parts once
\begin{align}
s\mathcal{L}_{p}\left(Z_{\left(+,\mathbf{k}\right)}Z_{\left(-,\mathbf{k}\right)}^{*}\right)-\left(Z_{\left(+,\mathbf{k}\right)}Z_{\left(-,\mathbf{k}\right)}^{*}\right)\left(t=0\right) & =\epsilon P_3 \\
 & -\epsilon\int_{0}^{\infty}\sum_{\beta,\gamma}V_{\left(+,\mathbf{k}\right)}^{\beta\gamma}\frac{d}{dt}\left(Z_{\beta}^{*}Z_{\gamma}^{*}Z_{\left(-,\mathbf{k}\right)}^{*}\right)\frac{e^{i\left(\omega_{\gamma}+\omega_{\beta}+\omega_{\left(+,\mathbf{k}\right)}\right)t}}{i\left(\omega_{\gamma}+\omega_{\beta}+\omega_{\left(+,\mathbf{k}\right)}\right)-s}e^{-st}dt\label{eq:anom1}\\\nonumber
 & -\epsilon\int_{0}^{\infty}\sum_{\beta,\gamma}V_{\left(-,\mathbf{k}\right)}^{\beta\gamma}\frac{d}{dt}\left(Z_{\left(+,\mathbf{k}\right)}Z_{\beta}Z_{\gamma}\right)\frac{e^{-i\left(\omega_{\gamma}+\omega_{\beta}+\omega_{\left(-,\mathbf{k}\right)}\right)t}}{i\left(\omega_{\gamma}+\omega_{\beta}+\omega_{\left(+,\mathbf{k}\right)}\right)-s}e^{-st}dt, \nonumber
\end{align}
where $P_3$ stands for third order polynomials which
do not contribute to the kinetic equation and $\mathcal{L}_{p}$ for
the Laplace transform. Let us consider one of the terms on the RHS 
\begin{align*}
 &\epsilon\int_{0}^{\infty}\frac{d}{dt}\left(Z_{\beta}^{*}Z_{\gamma}^{*}Z_{\left(-,\mathbf{k}\right)}^{*}\right)\frac{e^{i\left(\omega_{\gamma}+\omega_{\beta}+\omega_{\left(+,\mathbf{k}\right)}\right)t}}{i\left(\omega_{\gamma}+\omega_{\beta}+\omega_{\left(+,\mathbf{k}\right)}\right)-s}e^{-st}dt\\
 & =-\epsilon^{2}\int_{0}^{\infty}\left(\sum_{\beta',\gamma'}V_{\beta}^{\beta'\gamma'}Z_{\beta'}Z_{\gamma'}Z_{\gamma}^{*}Z_{\left(-,\mathbf{k}\right)}^{*}\right)\frac{d}{dt}\int_{0}^{t}\frac{e^{i\left(\omega_{\gamma}+\omega_{\beta}+\omega_{\left(+,\mathbf{k}\right)}\right)t'}}{i\left(\omega_{\gamma}+\omega_{\beta}+\omega_{\left(+,\mathbf{k}\right)}\right)-s}e^{-i\left(\omega_{\gamma'}+\omega_{\beta'}+\omega_{\beta}\right)t'}e^{-st'}dt+...
\end{align*}
Integrate by parts once more and take the average with respect to
the Gaussian initial distribution we arrive at
\begin{align}
\mathcal{L}_{p}\left(Z_{\left(+,\mathbf{k}\right)}Z_{\left(-,\mathbf{k}\right)}^{*}\right)-\frac{\left(Z_{\left(+,\mathbf{k}\right)}Z_{\left(-,\mathbf{k}\right)}^{*}\right)\left(t=0\right)}{s} & =-\epsilon^{2}\sum_{\beta,\gamma}V_{\left(+,\mathbf{k}\right)}^{\beta\gamma}\left(V_{\beta}^{\gamma\left(-,\mathbf{k}\right)}n_{\gamma}n_{\left(-,\mathbf{k}\right)}+\text{permutations}\right)\cross\\
 & \left(\frac{1}{s\left(i\left(\omega_{\gamma}+\omega_{\beta}+\omega_{\left(+,\mathbf{k}\right)}\right)-s\right)\left(2i\omega_{\left(+,\mathbf{k}\right)}-s\right)}\right)+...+O\left(\epsilon^{3}\right),
\end{align}
considering all the other contributions, take the inverse Laplace
transform and take the time derivative we obtain the equation for
the off diagonal correlator 
\begin{align}
\frac{d}{dt}\overline{Z_{\left(+,\mathbf{k}\right)}Z_{\left(-,\mathbf{k}\right)}^{*}} & =\epsilon^{2}\sum_{\beta,\gamma}V_{\left(+,\mathbf{k}\right)}^{\beta\gamma}\left(V_{\beta}^{\left(-,\mathbf{k}\right)\gamma}n_{\gamma}n_{\left(-,\mathbf{k}\right)}+V_{\gamma}^{\beta\left(-,\mathbf{k}\right)}n_{\beta}n_{\left(-,\mathbf{k}\right)}+V_{\left(-,\mathbf{k}\right)}^{\beta\gamma}n_{\beta}n_{\gamma}\right)\cross\\
 & e^{i\left(\omega_{\gamma}+\omega_{\beta}+\omega_{\left(+,\mathbf{k}\right)}\right)t}\int_{0}^{t}e^{-i\left(\omega_{\gamma}+\omega_{\left(-,\mathbf{k}\right)}+\omega_{\beta}\right)t'}dt'\\
 & +\epsilon^{2}\sum_{\beta,\gamma}V_{\left(-,\mathbf{k}\right)}^{\beta\gamma}\left(V_{\beta}^{\left(+,\mathbf{k}\right)\gamma}n_{\gamma}n_{\left(+,\mathbf{k}\right)}+V_{\gamma}^{\beta\left(+,\mathbf{k}\right)}n_{\beta}n_{\left(+,\mathbf{k}\right)}+V_{\left(+,\mathbf{k}\right)}^{\beta\gamma}n_{\beta}n_{\gamma}\right)\cross\nonumber \\
 & e^{-i\left(\omega_{\gamma}+\omega_{\beta}+\omega_{\left(-,\mathbf{k}\right)}\right)t}\int_{0}^{t}e^{i\left(\omega_{\gamma}+\omega_{\left(+,\mathbf{k}\right)}+\omega_{\beta}\right)t'}dt'.
\end{align}
Let us write explicitly the imaginary and real parts of one of the
oscillating terms on the right 

\begin{align}
\Im\left(e^{i\left(\omega_{\gamma}+\omega_{\beta}+\omega_{\left(+,\mathbf{k}\right)}\right)t}\int_{0}^{t}e^{-i\left(\omega_{\gamma}+\omega_{\left(-,\mathbf{k}\right)}+\omega_{\beta}\right)t}\right) & =i\cos\left(\omega_{\gamma}+\omega_{\beta}+\omega_{\left(+,\mathbf{k}\right)}\right)t\frac{1-\cos\left(\omega_{\gamma}+\omega_{\left(-,\mathbf{k}\right)}+\omega_{\beta}\right)t}{\left(\omega_{\gamma}+\omega_{\left(-,\mathbf{k}\right)}+\omega_{\beta}\right)}\\
 & -i\sin\left(\omega_{\gamma}+\omega_{\beta}+\omega_{\left(+,\mathbf{k}\right)}\right)t\frac{\sin\left(\omega_{\gamma}+\omega_{\left(-,\mathbf{k}\right)}+\omega_{\beta}\right)t}{\left(\omega_{\gamma}+\omega_{\left(-,\mathbf{k}\right)}+\omega_{\beta}\right)},\nonumber 
\end{align}

\begin{align}
\Re\left(e^{i\left(\omega_{\gamma}+\omega_{\beta}+\omega_{\left(+,\mathbf{k}\right)}\right)t}\int_{0}^{t}e^{-i\left(\omega_{\gamma}+\omega_{\left(-,\mathbf{k}\right)}+\omega_{\beta}\right)t}\right) & =-\sin\left(\omega_{\gamma}+\omega_{\beta}+\omega_{\left(+,\mathbf{k}\right)}\right)t\frac{1-\cos\left(\omega_{\gamma}+\omega_{\left(-,\mathbf{k}\right)}+\omega_{\beta}\right)t}{\left(\omega_{\gamma}+\omega_{\left(-,\mathbf{k}\right)}+\omega_{\beta}\right)}\\
 & -\cos\left(\omega_{\gamma}+\omega_{\beta}+\omega_{\left(+,\mathbf{k}\right)}\right)t\frac{\sin\left(\omega_{\gamma}+\omega_{\left(-,\mathbf{k}\right)}+\omega_{\beta}\right)t}{\left(\omega_{\gamma}+\omega_{\left(-,\mathbf{k}\right)}+\omega_{\beta}\right)}.\nonumber 
\end{align}
We will show that in the kinetic limit $\omega t\rightarrow\infty$
the function $f=\sin\left(\omega_{\gamma}+\omega_{\beta}+\omega_{\left(+,\mathbf{k}\right)}\right)t\frac{\sin\left(\omega_{\gamma}+\omega_{\left(-,\mathbf{k}\right)}+\omega_{\beta}\right)t}{\left(\omega_{\gamma}+\omega_{\left(-,\mathbf{k}\right)}+\omega_{\beta}\right)}$
converges weakly to $0$, this similarly follows for the rest of the
terms. First let us quickly show that for $x>0$ both functions $\sin\left(xs\right),t\sin\left(xs\right)$
converge weakly to $0$ as $s\rightarrow\infty$: Let $\phi$ be a
test function, then using integration by parts
\begin{align}
\lim_{t\rightarrow\infty}\int_{-\infty}^{\infty}\sin\left(xt\right)\phi\left(x\right)dx & =\lim_{t\rightarrow\infty}\int\frac{d}{dx}\left(-\frac{\cos\left(xt\right)}{t}\right)\phi\left(x\right)dx=\lim_{t\rightarrow\infty}\frac{1}{t}\int\cos\left(xt\right)\phi_{x}\left(x\right)dx=\lim_{t\rightarrow\infty}\frac{C}{t}=0\label{eq:sin}
\end{align}
and
\begin{align}
\lim_{t\rightarrow\infty}\int_{-\infty}^{\infty}t\sin\left(xt\right)\phi\left(x\right)dx & =\lim_{t\rightarrow\infty}\int\frac{d}{dx}\left(-\cos\left(xt\right)\right)\phi\left(x\right)dx=\lim_{t\rightarrow\infty}\int\frac{d}{dx}\left(\frac{\sin\left(xt\right)}{t}\right)\phi_{x}\left(x\right)dx\nonumber \\
 & =-\lim_{t\rightarrow\infty}\int\frac{\sin\left(xt\right)}{t}\phi_{xx}\left(x\right)dx=\lim_{t\rightarrow\infty}\frac{C}{t}=0.\label{eq:tsin}
\end{align}
Let $\Gamma\left(y\right)$ be a bounded domain , $g\left(y\right)$
a nice differential function and $y_{0}\in\left[-1,1\right]$ s.t
$\left(y-y_{0}\right)>0$ then 
\begin{align*}
\int_{\Gamma}dy\sin\left(\left(y-y_{0}\right)t\right)\frac{\sin\left(yt\right)}{y}g\left(y\right) & =\int dy\sin\left(\left(y-y_{0}\right)t\right)\frac{d}{dy}\int_{c}^{y}\frac{\sin\left(y't\right)}{y'}dy'g\left(y\right)\\
 & =-\int dy\frac{d}{dy}\left(\sin\left(\left(y-y_{0}\right)t\right)g\left(y\right)\right)\int_{c}^{y}\frac{\sin\left(y't\right)}{y'}dy'\\
 & =-\int dy\left(-t\cos\left(\left(y-y_{0}\right)t\right)g\left(y\right)+\sin\left(\left(y-y_{0}\right)t\right)g_{y}\left(y\right)\right)\int_{c}^{y}\frac{\sin\left(Ny't\right)}{y'}dy'
\end{align*}
taking the limit $t\rightarrow\infty$, using what we showed in (\ref{eq:sin},\ref{eq:tsin})
and $\left|\Theta\left(y\right)\right|=\left|\int_{c}^{y}\frac{\sin\left(Ny't\right)}{y'}dy'\right|<2$
we obtain 

\begin{align}
\lim_{t\rightarrow\infty}\int_{\Gamma}dy\sin\left(\left(y-y_{0}\right)t\right)\frac{\sin\left(yt\right)}{y}g\left(y\right) & \sim\int dy\lim_{t\rightarrow\infty}\left(-t\cos\left(\left(y-y_{0}\right)t\right)g\left(y\right)+\sin\left(\left(y-y_{0}\right)t\right)g_{y}\left(y\right)\right)2=0.\label{eq:sinsin}
\end{align}
Finally, we show that the integration of the collision kernel can
be brought to a form similar to (\ref{eq:sinsin}). Consider the integral
\begin{align}
\mathcal{I}_{\pm}= & \int K_{p}dK_{p}\int K_{q}dK_{q}\delta\left(\mathbf{k}\right)\int d\theta_{p}\int d\theta_{q}\sin\left(N\left(\cos\theta_{p}+\cos\theta_{q}-\cos\theta_{k}\right)t\right)\frac{\sin\left(N\left(\cos\theta_{p}+\cos\theta_{q}+\cos\theta_{k}\right)t\right)}{\left(\cos\theta_{p}+\cos\theta_{q}+\cos\theta_{k}\right)}\mathcal{K}\label{eq:col pm}
\end{align}
where $\mathcal{K}$ stands for terms in the collision kernel. Let
us change to the variables
\begin{align}
x & =\cos\theta_{p}-\cos\theta_{q}\\
y & =\cos\theta_{p}+\cos\theta_{q}+\cos\theta_{k}
\end{align}
Note that away from the resonant condition $y=0$ (\ref{eq:col pm})
is zero in the limit $\omega t\rightarrow\infty$. The the Jacobian
determinant of the coordinate transformation is $\det J=2\sin\theta_{p}\sin\theta_{q}$,
which is positive in the vicinity of $y=0$; take $\Gamma$ to be
a small vicinity of $y=0$ s.t $\det J\mid_{\Gamma}>0$ in the kinetic
time limit we are we are left with integral of the form 
\begin{align}
\lim_{\omega t\rightarrow\infty}\int K_{p}dK_{p}\int K_{q}dK_{q}\delta\left(\mathbf{k}\right)\lim_{t\rightarrow\infty}\int dx\int_{\Gamma}dy\det J\sin\left(N\left(y-2\cos\theta_{k}\right)t\right)\frac{\sin\left(Nyt\right)}{y}\mathcal{K}\left(x,y\right) & =0
\end{align}
this integral vanishes in the limit due to \eqref{eq:sinsin}.